\documentclass[aps,prd,reprint,nofootinbib,superscriptaddress,showkeys,floatfix,10pt]{revtex4-2}

\usepackage{amsmath,amssymb}
\usepackage{booktabs}
\usepackage{graphicx}
\graphicspath{{figures/}}
\makeatletter
\def\@bibdataout@aps{%
 \immediate\write\@bibdataout{@CONTROL{apsrev42Control,author="9",editor="0",pages="0",title="0",year="1"}}%
 \if@filesw
  \immediate\write\@auxout{\string\citation{apsrev42Control}}%
 \fi
}
\def\@fltstk{}
\def\ltxgrid@warn#1{}
\makeatother

\begin{document}

\title{Long-lived ringing of Bronnikov--Kim brane-world black holes and wormholes}
\author{Milena Skvortsova}
\email{milenas577@mail.ru}
\affiliation{Institute of Gravitation and Cosmology, Peoples' Friendship University of Russia (RUDN University), 6 Miklukho-Maklaya Street, Moscow 117198, Russia}
\date{\today}

\begin{abstract}

Massive-field perturbations on black-hole and wormhole
backgrounds can form long-lived quasinormal ringing; this behaviour is not
confined to brane-world spacetimes, but it is also not universal across
compact-object geometries.
Recent work on a brane-world black hole reported an unusual limiting behaviour in
which the real oscillation frequency tends to zero as the scalar mass grows. We
test how general this phenomenon is in the Bronnikov--Kim effective brane-world
geometries, treating both the black-hole and wormhole sides of the same families.
The comparison uses two complementary numerical treatments: Leaver's
continued-fraction method for the black-hole side, after reducing the wave equation for the second Bronnikov--Kim family
to rational form, and Chebyshev pseudospectral collocation for
regular wormhole throats. The
black hole of the second Bronnikov--Kim family shows the standard
quasi-resonant pattern: the damping rates of the scalar branches considered here
become very small, but the real frequencies remain finite.  Representative wormholes from the first and second
Bronnikov--Kim families also develop increasingly long-lived massive scalar
modes, and extending the
second-family wormhole scan to larger field mass reduces the damping by more
than a factor of four relative to the massless case. Thus the robust brane-world effect found here is
the appearance of long-lived massive-field ringing in both black holes and
wormholes, whereas the vanishing of the real frequency is metric- or
branch-dependent.
\end{abstract}


\maketitle

\section{Introduction}

Quasinormal modes (QNMs) are the characteristic damped oscillations selected by
boundary conditions at compact objects and at infinity. They are central both to
black-hole spectroscopy and to the mathematical theory of wave propagation on
curved backgrounds \cite{Kokkotas:1999bd,Berti:2009kk,Konoplya:2011qq,Bolokhov:2025rng}. The mass of a
field changes this spectral problem more deeply than by a small shift of the
potential peak. In an asymptotically flat geometry the potential no longer
vanishes at infinity but approaches $\mu^2$, so the outgoing wave number becomes
$k=\sqrt{\omega^2-\mu^2}$ and the spectrum is tied to the global interpolation
between the compact object and the massive asymptotic plateau. The late-time
response is also reorganized: instead of the usual massless power-law tail one
finds oscillatory massive tails whose decay depends on the mass scale and on the
large-distance structure of the spacetime~\cite{Koyama:2000hj,Moderski:2001tk,Koyama:2001qw,Jing:2004zb,Koyama:2001ee,Rogatko:2007zz,Gibbons:2008gg,Gibbons:2008rs,Dubinsky:2024jqi}.

This global sensitivity is the reason massive fields are useful probes of
non-Schwarzschild compact objects. The best-known spectral manifestation is the
formation of quasi-resonances, where the damping can be made arbitrarily small
for selected values of the field mass~\cite{Konoplya:2004wg,Ohashi:2004wr}.
Long-lived branches and other mass-induced deformations of the QNM spectrum have
been found for different spins and in a wide range of black-hole or
compact-object backgrounds~\cite{Konoplya:2018qov,Churilova:2020bql,Zhidenko:2006rs,Burikham:2017gdm,Aragon:2020teq,Konoplya:2017tvu,Fernandes:2021qvr,Percival:2020skc,Gonzalez:2022upu,Zinhailo:2018ska,Skvortsova:2026jtx,Bolokhov:2026dfg}.
The same phenomenon, with geometry-dependent variations, has also been explored
in regular, deformed, and modified-gravity spacetimes~\cite{Konoplya:2007zx,Lutfuoglu:2025hwh,Skvortsova:2024eqi,Lutfuoglu:2025eik,Skvortsova:2025cah,Lutfuoglu:2025kqp,Lutfuoglu:2025hjy,Lutfuoglu:2026fpx,Skvortsova:2026unq,Bolokhov:2023ruj,Bolokhov:2026dzn,Bolokhov:2026uol}.
At the same time, quasi-resonant behaviour is not guaranteed in all cases: some backgrounds or
perturbation types do not display the same approach to vanishing damping, or
do so only on particular branches~\cite{Konoplya:2005hr,Zinhailo:2024jzt}.
Thus a massive scalar field tests not merely the height of a potential barrier,
but the way a specific geometry couples the near-peak region to the asymptotic
mass scale.

Brane-world gravity provides a particularly natural arena in which to ask how
robust these massive-field effects are. In Randall--Sundrum and Dvali--Gabadadze--Porrati-type
scenarios~\cite{Randall:1999ee,Randall:1999vf,Dvali:2000hr}, the observed
universe is a four-dimensional brane embedded in a higher-dimensional bulk. The
standard-model fields are usually confined to the brane, while gravity can probe
the extra dimensions, so the gravitational field induced on the brane need not
obey the ordinary four-dimensional vacuum Einstein equations. Such models were
motivated in part by ideas from string theory and by attempts to reformulate the
hierarchy problem, late-time cosmic acceleration, and other puzzles of physics
beyond the standard model. For compact objects, the important consequence is
more concrete: bulk curvature and projected Weyl stresses can modify the
four-dimensional metric and can even allow static black-hole and wormhole
geometries that have no direct Schwarzschild analogue.

Quasinormal ringing has therefore become one of the standard probes of higher-dimensional
and brane-world gravity. A broad set of black-hole QNM calculations in such
settings can be found in
Refs.~\cite{2753763,Sakalli:2022swm,Churilova:2021tgn,daRocha:2017lqj,Soleimani:2016mfh,Yang:2014cra,Bigazzi:2013jqa,Kaminski:2009dh,Zhidenko:2009zx,Chen:2007jz,Kanti:2006ua,Abdalla:2007zz,BendasoliPavan:2006vk,Abdalla:2006qj,Kodama:2009rq,Konoplya:2013sba,Konoplya:2008rq,Konoplya:2007jv,Kodama:2009bf,Konoplya:2017zwo}.
The related problem of fields localized on the brane has also been studied in
several representative backgrounds~\cite{Zinhailo:2024jzt,Zhidenko:2008fp,Kanti:2005xa,Chung:2015mna,Seahra:2005us,Koyama:2005gh,Seahra:2005wk}.
These works show that the ringdown spectrum is a sensitive diagnostic of the
bulk-induced corrections because it depends both on the near-peak structure of
the effective potential and on the asymptotic boundary conditions.

In a brane-world setting this sensitivity to both the asymptotic plateau and the
interior geometry is especially useful. The Casadio--Fabbri--Mazzacurati (CFM)
solution~\cite{Casadio:2001jg}, originally proposed as a brane-world black-hole
metric and later examined near its black-hole--wormhole transition through QN
frequencies~\cite{Abdalla:2006qj}, illustrates the point. A recent analysis of a massive
scalar field perturbations in this background found a particularly strong branch-dependent
effect: as the field mass grows, one mode can lose its real oscillation
frequency and be replaced by an overtone~\cite{Lutfuoglu:2026cfm}. Because
massive spectra are known to be highly geometry dependent, this behaviour should
not be assumed to follow from the brane-world label alone. Bronnikov and Kim
constructed broader effective brane-world families containing both black-hole and
wormhole branches~\cite{Bronnikov:2002rn}. They therefore provide a natural test
of two separate issues: whether the CFM zero-real-frequency behaviour is generic,
and whether the more robust feature is instead the formation of long-lived
massive modes when Weyl-induced deformations compete with the massive
asymptotic plateau.

We address these questions for the Bronnikov--Kim effective brane-world
geometries, which originate in the gravitational equations induced on a
four-dimensional brane embedded in a higher-dimensional
bulk~\cite{Bronnikov:2002rn,Bronnikov:2019sbx}. These geometries are well
suited for the comparison because the same families include black-hole and
wormhole branches. We will show that the monopole fundamental mode, the first three monopole overtones, and the dipole
fundamental mode all develop small damping rates while their real parts remain
finite. On the wormhole side we use pseudospectral collocation for
representative Bronnikov--Kim--1 and Bronnikov--Kim--2 solutions. The even-symmetry-sector fundamental mode
shows the same qualitative trend: increasing the field mass lowers the damping
rate without driving the oscillation frequency to zero.

Our numerical treatment uses two complementary methods. For the black-hole
case we first transform the Bronnikov--Kim--2 scalar wave equation to a
rational form and then solve the spectral problem by Leaver's
continued-fraction method. For the wormhole side we use a one-exterior symmetry
reduction of the smooth two-ended problem and solve the resulting eigenvalue
problem by Chebyshev pseudospectral collocation. The quasi-resonant regime is
nevertheless numerically delicate.  The known ingoing or
outgoing asymptotic wave behaviour
must be separated off before a Frobenius recurrence or pseudospectral matrix is
formed; for a symmetric wormhole, any half-domain
matching condition is only a smooth matching condition. Otherwise the numerical
problem represents the wrong asymptotic conditions. The paper is
organized as follows. Section~II gives the massive scalar radial equation,
Sec.~III summarizes the Bronnikov--Kim black-hole and wormhole geometries,
Sec.~IV derives the continued fraction for the black-hole's spectral problem, and
Sec.~V presents the black-hole and wormhole spectra.
The same section also gives the eikonal Wentzel--Kramers--Brillouin
(WKB) limit for the single-peak black-hole barriers, providing an analytic
large-$\ell$ check for the BK--1 and BK--2 geometries.

\section{Scalar perturbation equation}

The backgrounds considered below are four-dimensional metrics induced on a
brane embedded in a higher-dimensional bulk. In the effective brane-world
description used for the Bronnikov--Kim solutions~\cite{Shiromizu:1999wj,Bronnikov:2002rn,Bronnikov:2019sbx},
the induced metric $g_{\mu\nu}$ satisfies
\begin{align}
 {}^{(4)}G_{\mu\nu}
 &=-\Lambda_4 g_{\mu\nu}+\kappa_4^2 T_{\mu\nu}
 +\kappa_5^4 S_{\mu\nu}-{\cal E}_{\mu\nu},\notag\\
 {\cal E}_{\mu\nu}
 &={}^{(5)}C_{ABCD}n^A n^C e^B{}_{\mu}e^D{}_{\nu}.
 \label{eq:brane-effective-equations}
\end{align}
Here $S_{\mu\nu}$ is the local term quadratic in the brane stress tensor,
whereas ${\cal E}_{\mu\nu}$ is the projection of the five-dimensional Weyl tensor
and carries the nonlocal tidal influence of the bulk. The Bronnikov--Kim
solutions used in this work belong to the vacuum brane case with
$T_{\mu\nu}=0$ and $\Lambda_4=0$. The effective equations then reduce to
\begin{equation}
 R_{\mu\nu}=-{\cal E}_{\mu\nu},
 \qquad
 {\cal E}^{\mu}{}_{\mu}=0,
 \qquad
 \nabla^{\mu}{\cal E}_{\mu\nu}=0,
 \qquad
 R=0.
 \label{eq:brane-vacuum}
\end{equation}
Thus the induced metric is not required to be a vacuum solution of ordinary
four-dimensional general relativity; it is constrained instead by the traceless
Weyl source generated by the bulk. We study a massive scalar field
in the background of this effective four-dimensional geometry.

We use the static, spherically symmetric line element
\begin{equation}
 ds^2=-A(r)dt^2+\frac{dr^2}{B(r)}+r^2d\Omega^2,
\end{equation}
The massive scalar field obeys
\begin{equation}
 \left(\nabla^\alpha\nabla_\alpha-\mu^2\right)\Phi=0.
\end{equation}
After separation of variables,
\begin{equation}
 \Phi=e^{-i\omega t}Y_{\ell m}(\theta,\phi)\frac{\psi(r)}{r},
\end{equation}
the radial equation takes the Schr\"odinger form (see, for instance, \cite{Carter:1968ks,Konoplya:2018arm})
\begin{equation}
 \frac{d^2\psi}{dr_*^2}+\left[\omega^2-V(r)\right]\psi=0,
 \qquad
 \frac{dr_*}{dr}=\frac{1}{\sqrt{A(r)B(r)}}.
 \label{eq:schrodinger}
\end{equation}
The effective potential is
\begin{equation}
 V(r)=A(r)\left(\frac{\ell(\ell+1)}{r^2}+\mu^2\right)
    +\frac{A'(r)B(r)+A(r)B'(r)}{2r}.
 \label{eq:general-potential}
\end{equation}
Here and below a prime denotes $d/dr$, while a subscript after a comma denotes
differentiation with respect to the indicated variable. The integer $\ell$ is
the spherical-harmonic index, $\mu$ is the scalar-field mass, and $n=0,1,2,\ldots$
labels the overtone number, with $n=0$ denoting the fundamental branch.

Equivalently, in the areal coordinate,
\begin{equation}
 A(r)B(r)\psi''+\frac{\left[A(r)B(r)\right]'}{2}\psi'
 +\left[\omega^2-V(r)\right]\psi=0.
 \label{eq:r-equation}
\end{equation}
Equations~\eqref{eq:schrodinger}--\eqref{eq:r-equation} fix the normalization
and sign conventions used below: stable QNMs have
$\operatorname{Im}\omega<0$ with the time dependence $e^{-i\omega t}$.

\begin{figure*}[tp]
\centering
\includegraphics[width=0.92\textwidth]{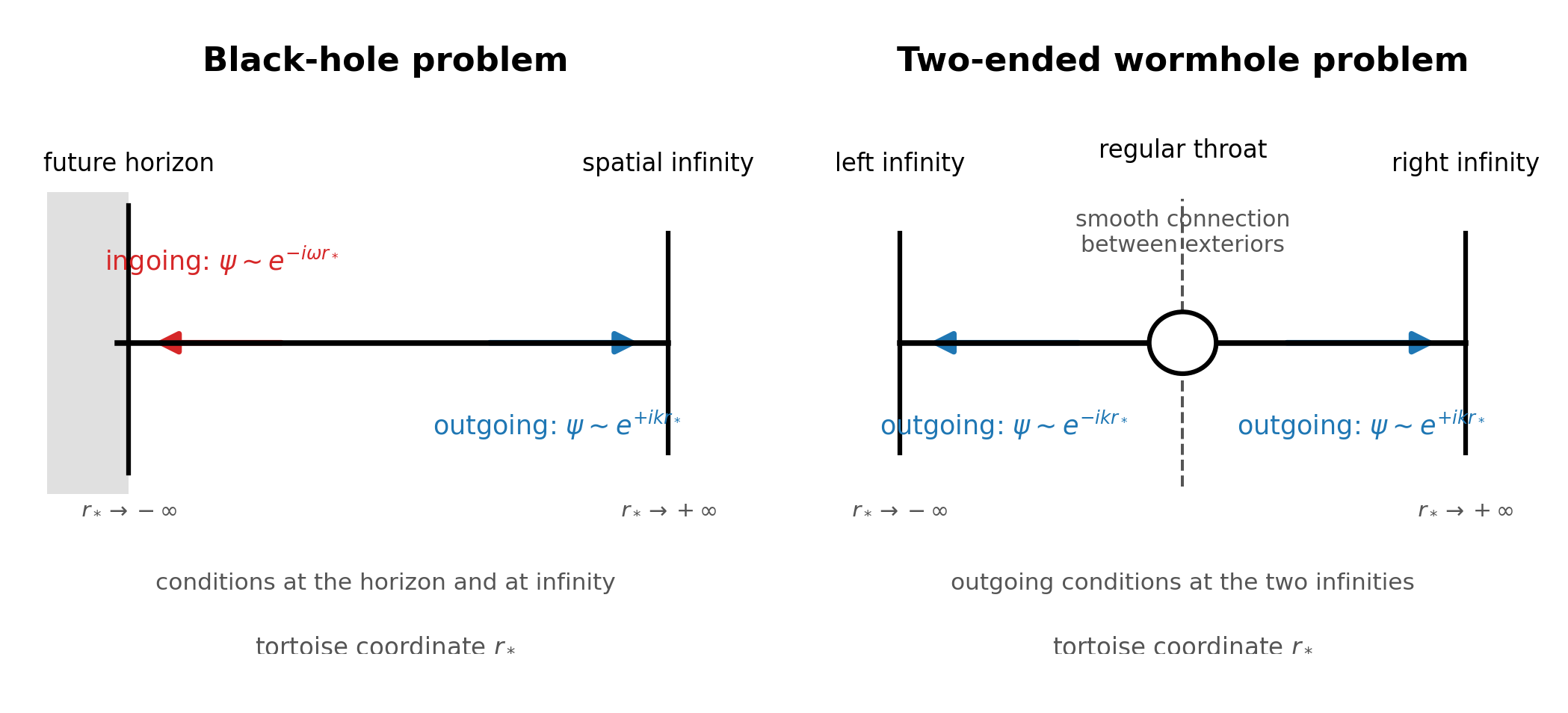}
\caption{Schematic boundary conditions for the two radial problems considered in
this work. Black-hole QNMs are ingoing at the horizon and outgoing at spatial
infinity. Wormhole resonances are outgoing at both asymptotic ends; for a symmetric wormhole one may use a
reflection-symmetry sector to reduce the full two-ended problem to a half-line,
but the throat remains a regular interior point, not a boundary}
\label{fig:boundary-conditions}
\end{figure*}

The quasinormal boundary conditions depend on the asymptotic endpoints of the radial
problem. For black holes one requires a purely ingoing
wave at the future horizon and a purely outgoing massive wave at
infinity. If the horizon is a simple zero of $A B$, the local solution has the
form
\begin{equation}
 \psi\propto e^{-i\omega r_*},\qquad r\to r_h.
\end{equation}
At the asymptotically flat end the potential tends to $\mu^2$, and the outgoing
black-hole solution is
\begin{equation}
 \psi\propto e^{+ik r_*},\qquad
 k=\sqrt{\omega^2-\mu^2},\qquad r_*=+\infty.
\end{equation}
For a two-ended wormhole there is no ingoing horizon condition. Instead the
resonance condition is outgoing radiation at both asymptotic ends,
\begin{equation*}
 \psi\propto e^{-ik r_*},\quad r_*\to-\infty,
 \qquad
 \psi\propto e^{+ik r_*},\quad r_*\to+\infty.
\end{equation*}
The throat is a regular interior point of the two-ended tortoise-coordinate problem, not a
physical boundary. Thus the scalar field is continued smoothly through the
throat, and no reflecting or absorbing condition is imposed there in the physical
problem. For reflection-symmetric examples, one may later use a
symmetry-sector decomposition as a computational reduction of the full two-ended
problem; this is only a
matching of the two sides, not an infinite barrier at the throat. The square
root in $k$ is taken on the branch continuously connected to $k=\omega$ in the
massless limit. This branch choice is important when the modes approach a
quasi-resonant endpoint, because $k$ then becomes small and the outgoing
condition is the quantity that selects the discrete spectrum. The black-hole
boundary conditions, the outgoing wormhole conditions at the two asymptotic ends,
and the position of the wormhole throat are summarized schematically in
Fig.~\ref{fig:boundary-conditions}.

The effective potentials shown in Fig.~\ref{fig:bk-effective-potentials} are positive definite. Hence the self-adjoint operator
\begin{equation}
\mathcal{D} = -\frac{\partial^2}{\partial r_*^2} + V(r)
\end{equation}
is positive in the Hilbert space of square-integrable functions of the tortoise coordinate. All solutions
of the wave equation corresponding to compact-support initial data are therefore bounded \cite{Abdalla:2006qj,Konoplya:2025ect}. Thus, the quasinormal spectrum does not allow growing modes leading to instability.

\section{Bronnikov--Kim geometries}
\label{sec:geometries}

The Bronnikov--Kim construction supplies explicit static solutions of the
vacuum brane equations~\eqref{eq:brane-vacuum}. Since only the traceless Weyl
source is retained, the condition $R=0$ replaces the ordinary four-dimensional
vacuum Einstein equations. This weaker condition permits non-Schwarzschild
black-hole and wormhole geometries, with the detailed bulk geometry left
unspecified in the effective treatment~\cite{Bronnikov:2002rn}.

Bronnikov and Kim derived simple static metrics that interpolate between
black-hole and wormhole behaviour in this effective brane-world setting. The
first Bronnikov--Kim family is best introduced before any numerical choice of units. With mass scale $M$ and
branch parameter $r_0$ the metric functions can be written as

\begin{align}
 A(r)=A_1(r)&=\left(1-\frac{2M}{r}\right)^2,\notag\\
B(r)= B_1(r)&=\left(1-\frac{r_0}{r}\right)
     \left(1-\frac{r_1}{r}\right),
 \quad r_1=\frac{M r_0}{r_0-M}.
 \label{eq:bk1}
\end{align}

For $r_0>2M$ this metric describes a wormhole with a regular
throat at $r=r_0$, where $B_1$ vanishes while $A_1$ stays finite. For $r_0<2M$ we have a black hole geometry. The extremal
black-hole case is obtained only as the special case $r_0=2M$, for which
$r_1=2M$ and

\begin{equation}
 A_1(r)=B_1(r)=\left(1-\frac{2M}{r}\right)^2.
 \label{eq:bk1-extremal}
\end{equation}

The normalized form $A_1=B_1=(1-1/R)^2$ follows after the
coordinate rescaling $R=r/(2M)$; it should not be used as the definition of the
family.  The numerical
normalization and representative parameter values are stated later, together
with the tables.
For the extremal case, the dimensionless tortoise coordinate is
\begin{equation}
 R_*=R+2\ln(R-1)-\frac{1}{R-1},
\end{equation}
so, with the dimensionless frequency $\widehat\omega=2M\omega$, the ingoing
factor contains an irregular exponential,

\begin{equation}
 e^{-i\widehat\omega R_*}\propto
 \exp\left(\frac{i\widehat\omega}{x}\right)x^{-2i\widehat\omega},
 \qquad x=\frac{R-1}{R}.
 \label{eq:bk1-horizon-factor}
\end{equation}

This observation is used only to distinguish the
extremal BK--1 case from the non-extremal BK--2 continued-fraction problem
studied below; it is not the starting point for defining the Bronnikov--Kim
families.
The BK--1 wormhole used in the numerical section is therefore an
independent two-ended representative of the same first family, not an extremal
black-hole limit.

The second Bronnikov--Kim
family is likewise written first with its length scale explicit \cite{Bronnikov:2003gx}:

\begin{align}
 A_2(r)&=1-\frac{a^2}{r^2},\notag\\
 B_2(r)&=A_2(r)\left(1+\frac{C-1}{\sqrt{2r^2/a^2-1}}\right).
 \label{eq:bk2}
\end{align}

Here $a$ is the radius of the $C>0$ black-hole horizon. Introducing the dimensionless
radius $\rho=r/a$ gives the form used in the algebra below,

\begin{equation}
 A_2(\rho)=1-\frac{1}{\rho^2},
 \quad
 B_2(\rho)=A_2(\rho)\left(1+\frac{C-1}{\sqrt{2\rho^2-1}}\right).
\end{equation}

Thus using the dimensionless coordinate is a rescaling, not an early numerical
choice of the horizon radius.
For black holes in this parametrization one has $C>0$.
For $C<0$ the same metric describes a two-sided wormhole; the zero of $B_2$ is
then the throat rather than an event horizon.  The numerical
choice of $C$ is stated later together with the wormhole table.
Indeed, in the $y$ coordinate (see eq. (\ref{19})) $B_2$ is proportional to $y+C-1$, so the regular
throat of the $C<0$ branch is located at $y_{\rm th}=1-C$.
The case $C=1$ reduces to

\begin{equation}
 A_2(r)=B_2(r)=1-\frac{a^2}{r^2},
 \quad
 A_2(\rho)=B_2(\rho)=1-\frac{1}{\rho^2}.
 \label{eq:bk2-c1}
\end{equation}

For $C\ne1$ the square root in Eq.~\eqref{eq:bk2} is only apparent.
In the rationalized formulae below, we denote the dimensionless combination $\rho\equiv r/a$ as $r$ for simplicity. With
\begin{equation}\label{19}
 y=\sqrt{2r^2-1},
 \qquad
 r^2=\frac{y^2+1}{2},
\end{equation}
the metric functions become rational,
\begin{equation}
 A_2(y)=\frac{y^2-1}{y^2+1},
 \qquad
 B_2(y)=A_2(y)\frac{y+C-1}{y}.
 \label{eq:bk2-y-metric}
\end{equation}
The derivative conversion used in the radial equation is
\begin{equation}
 \frac{dy}{dr}=\frac{2r}{y}=\frac{\sqrt{2(y^2+1)}}{y},
 \qquad
 \frac{d^2y}{dr^2}=-\frac{2}{y^3}.
 \label{eq:y-derivatives}
\end{equation}
Substituting these relations into Eq.~\eqref{eq:r-equation} gives the
$\psi_{yy}$ and $\psi_y$ coefficients displayed in
Eq.~\eqref{eq:bk2-y-equation}.
Putting $P(y)=A_2(y)B_2(y)$, Eq.~\eqref{eq:r-equation} becomes
\begin{align}
 &\frac{2P(y)(y^2+1)}{y^2}\psi_{yy}
 +\left[\frac{(y^2+1)P_y(y)}{y^2}
 -\frac{2P(y)}{y^3}\right]\psi_y\notag\\
 &\hspace{1.0cm}
 +\left[\omega^2-V(y)\right]\psi=0,
 \label{eq:bk2-y-equation}
\end{align}
where
\begin{align}
 V(y)&=A_2(y)\left(\frac{2\ell(\ell+1)}{y^2+1}+\mu^2\right)\notag\\
 &\quad
 +\frac{A_{2,y}(y)B_2(y)+A_2(y)B_{2,y}(y)}{y}.
 \label{eq:bk2-y-potential}
\end{align}
Equivalently, the potential can be written explicitly as the rational function
\begin{align}
 V(y)&=\frac{y^2-1}{y^2+1}
 \biggl[\frac{2\ell(\ell+1)}{y^2+1}+\mu^2\biggr]\notag\\
 &\quad+\frac{8(y^2-1)(y+C-1)}{y(y^2+1)^3}
 -\frac{(C-1)(y^2-1)^2}{y^3(y^2+1)^2}.
 \label{eq:bk2-y-potential-explicit}
\end{align}
Thus the square-root metric admits an exact rational radial equation not only at
isolated parameter values, but throughout the positive-$C$ black-hole range.
Any truncation of $\sqrt{2r^2-1}$ is therefore a matter of numerical convenience
rather than necessity.

For generic $C$ it is useful to compactify the rationalized coordinate by
\begin{equation}
 z=1-\frac{1}{y},\qquad y=\frac{1}{1-z},
\end{equation}
so that the horizon is at $z=0$ and infinity at $z=1$. In this coordinate
\begin{align}
 A_2(z)&=\frac{z(2-z)}{1+(1-z)^2},\notag\\
 B_2(z)&=A_2(z)\left[C-(C-1)z\right].
 \label{eq:bk2-z-metric}
\end{align}
and the wave equation again has only rational coefficients. The endpoint asymptotics are
\begin{align}
 r_*&=\frac{1}{2\sqrt{C}}\ln z+\mathcal{O}(1),\qquad z\to0,\notag\\
 r_*&=\frac{1}{\sqrt{2}(1-z)}+\frac{C-1}{2\sqrt{2}}\ln(1-z)+\mathcal{O}(1),
 \qquad z\to1.
 \label{eq:general-c-tortoise-asymptotics}
\end{align}
These asymptotic formulae identify the singular wave behaviour
at the two ends of the compact interval. The product of these known endpoint
behaviours will be referred to as the asymptotic boundary factor, or endpoint
factor. For a black hole it is the ingoing wave at the horizon multiplied by
the outgoing massive wave at infinity, both rewritten in the compact coordinate.
The factorization $\psi=F u$, with this known factor $F$, imposes the
quasinormal boundary conditions analytically and leaves a residual function $u$
that is regular enough to expand or collocate. Thus ``removing'' the factor means substituting
$\psi=F u$ before forming the recurrence or matrix; it does not discard any
physical boundary condition.
Thus the natural generalization of the Leaver factor is
\begin{align}
 \psi(z)&=z^{-i\omega/(2\sqrt{C})}
 (1-z)^{i k(C-1)/(2\sqrt{2})}\notag\\
 &\quad\times \exp\left[\frac{i k}{\sqrt{2}(1-z)}\right]u(z).
 \label{eq:general-c-leaver-factor}
\end{align}
with $k=\sqrt{\omega^2-\mu^2}$. After this factor is removed, $u(z)$ satisfies
an equation with rational coefficients, and multiplying by a common denominator
again gives a finite-band recurrence.

The practical difference between positive values of $C$ is
the location of the additional zero of $B_2$. In the $z$ plane it sits at
\begin{equation}
 z_C=\frac{C}{C-1},
\end{equation}
with the $C=1$ case understood as the limit in which this zero is absent. For
$C>1$ this point lies beyond infinity, and for $1/2<C<1$ it lies to the left of
$z=-1$; in both cases it is outside the unit disk of the horizon expansion. At
$C=1/2$ it lies on the boundary, while for $0<C<1/2$ it lies inside the unit
circle and a single Frobenius series in the coordinate $z$ cannot reach
infinity. Those smaller positive values can still be treated either by the
midpoint analytic-continuation strategy described below or by changing the
compact coordinate. A useful choice is
\begin{equation}
 x=\frac{y-1}{y-y_0},\qquad y_0=\max(0,1-C),
 \qquad y=\frac{1-y_0 x}{1-x}.
 \label{eq:bk2-x-arbitrary-c}
\end{equation}
It reduces to $z=1-1/y$ for $C\ge1$, while for $0<C<1$ it sends the additional
zero $y=1-C$ to $x=\infty$ and keeps the physical interval at $0\le x\le1$.
The asymptotic boundary factor in this coordinate is the same as
Eq.~\eqref{eq:general-c-leaver-factor} except that the irregular exponential at
infinity is multiplied by $1-y_0$,
\begin{align}
 \psi(x)&=x^{-i\omega/(2\sqrt{C})}
 (1-x)^{i k(C-1)/(2\sqrt{2})}\notag\\
 &\quad\times \exp\left[\frac{i k(1-y_0)}{\sqrt{2}(1-x)}\right]u(x).
 \label{eq:general-c-x-leaver-factor}
\end{align}
The case $C=1$ is therefore not the only rational case; it is the special case
with the shortest asymptotic
boundary factor and no extra finite singularity near the interval. For $C=0$
the near-horizon zero is degenerate, while $C<0$ belongs to the wormhole side
of the black-hole--wormhole transition. Those cases require boundary conditions different from the
black-hole ingoing condition. We handle them below with a wormhole pseudospectral
calculation.

\begin{figure*}[tp]
\centering
\includegraphics[width=\textwidth]{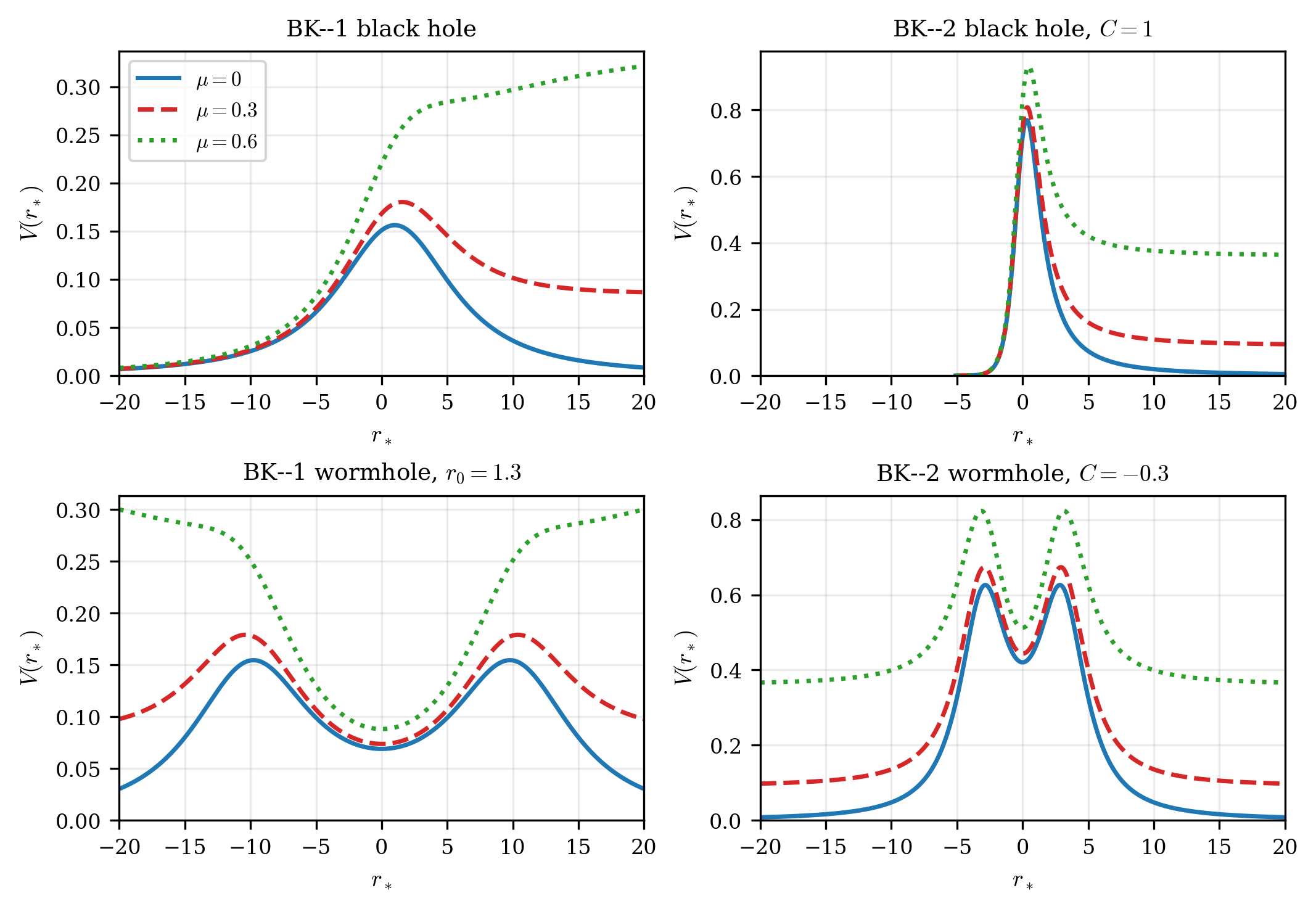}
\caption{Representative effective potentials as functions of the tortoise
coordinate for Bronnikov--Kim black holes and wormholes. The curves correspond
to scalar dipoles with $\ell=1$ and $\mu=0$, $0.3$, and $0.6$, as indicated in
the legend. The black-hole panels use the BK--1 extremal horizon and the BK--2 black hole, while the wormhole panels use the representative throats
employed later in the mode calculations.}
\label{fig:bk-effective-potentials}
\end{figure*}

Before turning to the QNM calculations, it is useful to compare the shapes of
representative scalar effective potentials in the tortoise coordinate. Figure~\ref{fig:bk-effective-potentials}
shows the dipole potential, $\ell=1$, for $\mu=0$, $0.3$, and $0.6$, for one
black-hole and one wormhole example from each Bronnikov--Kim family. For black
holes, $r_*$ runs from the horizon to spatial infinity. For wormholes, the
throat is placed at $r_*=0$, and the second exterior is shown by reflecting the
same one-sided potential. The massive curves illustrate the rise of the
large-distance plateau in Eq.~\ref{eq:general-potential}.

\section{Leaver method for the Bronnikov--Kim--2 metric}
\label{sec:leaver}

Leaver method \cite{Leaver:1986gd,Leaver:1990zz} is based on the expansion of the solution of the wave equation with quasinormal boundary conditions into the Frobenius series. As it is based on the convergent procedure, it was used in numerous works for precise finding of quasinormal modes \cite{Onozawa:1996ux,Konoplya:2004uk,Bolokhov:2023bwm,Lutfuoglu:2025ljm,Konoplya:2020hyk,Dias:2021yju,Konoplya:2023ahd,Chung:2023zdq,Bolokhov:2024bke,Lutfuoglu:2025qkt,Konoplya:2023ppx,Casals:2011aa,Lutfuoglu:2025bsf,Konoplya:2023aph,Saka:2025xxl,Benda:2025tni}. In order to apply the Frobenius expansion, the coefficient of the wave equation must have rational coefficients.

For $C=1$ define the compact coordinate
\begin{equation}
 x=1-\frac{1}{r},
 \qquad s=1-x,
 \qquad f(x)=1-\frac{1}{r^2}=x(2-x).
\end{equation}
The event horizon is at $x=0$ and spatial infinity at $x=1$. The tortoise
coordinate is elementary,
\begin{equation}
 r_*=r+\frac12\ln\frac{r-1}{r+1},
\end{equation}
and the potential reduces to
\begin{equation}
 V(r)=f(r)\left[\frac{\ell(\ell+1)}{r^2}+\mu^2+\frac{2}{r^4}\right].
\end{equation}
In the compact coordinate the scalar equation can be written as
\begin{equation}
 a_2(x)\psi_{xx}+a_1(x)\psi_x+a_0(x)\psi=0,
 \label{eq:c1-x-equation}
\end{equation}
with polynomial coefficients
\begin{align}
 a_2(x)&=f(x)^2s^4,\\
 a_1(x)&=2f(x)s^5-2f(x)^2s^3,\\
 a_0(x)&=\omega^2-f(x)\left[\ell(\ell+1)s^2+\mu^2+2s^4\right].
\end{align}
Near the horizon, $r_*\sim \frac12\ln x$, hence an ingoing wave behaves as
$x^{-i\omega/2}$. At infinity, $r_*=r+O(r^{-1})$ with $r=(1-x)^{-1}$, and the
outgoing massive wave contains $\exp[ik/(1-x)]$. We therefore factor
\begin{align}
 \psi(x)&=x^{-i\omega/2}\exp\left(\frac{ik}{1-x}\right)u(x),\notag\\
 u(x)&=\sum_{n=0}^{\infty}a_n x^n,
 \qquad a_0=1.
 \label{eq:leaver-factor-c1}
\end{align}
If $g(x)$ denotes the logarithmic derivative of the
asymptotic boundary factor multiplying $u$ in Eq.~\eqref{eq:leaver-factor-c1},
\begin{equation}
 g(x)=-\frac{i\omega}{2x}+\frac{ik}{(1-x)^2},
\end{equation}
then $u(x)$ satisfies
\begin{equation}
 b_2(x)u''+b_1(x)u'+b_0(x)u=0,
 \label{eq:regular-u-equation}
\end{equation}
where
\begin{align}
 b_2&=a_2,\\
 b_1&=a_1+2a_2g,\\
 b_0&=a_0+a_1g+a_2\left(g'+g^2\right).
\end{align}
After multiplying by the common denominator and canceling common algebraic factors in the
three coefficients, the $b_j$ are polynomials and may be expanded as
\begin{equation}
 b_j(x)=\sum_{m=0}^{d_j}b_{j,m}x^m.
\end{equation}
Substitution of the Frobenius series gives a finite-band recurrence. The
coefficient of $x^N$ is
\begin{align}
 &\sum_m b_{0,m}a_{N-m}
 +\sum_m b_{1,m}(N-m+1)a_{N-m+1}\notag\\
 &\quad
 +\sum_m b_{2,m}(N-m+2)(N-m+1)a_{N-m+2}=0,
 \label{eq:wide-recurrence}
\end{align}
with $a_n=0$ for $n<0$. This finite-band recurrence is reduced to a three-term
recurrence by Gaussian elimination,
\begin{equation}
 \alpha_n a_{n+1}+\beta_n a_n+\gamma_n a_{n-1}=0.
 \label{eq:three-term}
\end{equation}
The minimal-solution condition is imposed by the continued fraction
\begin{equation}
 \beta_0-\frac{\alpha_0\gamma_1}{\beta_1-}
 \frac{\alpha_1\gamma_2}{\beta_2-}
 \frac{\alpha_2\gamma_3}{\beta_3-}\cdots=0.
 \label{eq:cf-fundamental}
\end{equation}
The $n$th overtone is obtained by the standard inverted continued fraction,
where the recurrence is solved both inward from large index and outward from the
horizon and then matched at the inversion number. Numerically, the tail of the
right continued fraction is seeded by the lower-half-plane root of the limiting
quadratic for $a_{N+1}/a_N$; this stabilizes the nearly real quasi-resonant
regime. Roots were followed by continuation in $\mu$, using the solution at one
mass as the initial guess for the next mass. Near the endpoint the truncation
index was increased until the quoted digits were stable at the level needed to
identify whether $\operatorname{Re}\omega$ was tending to zero.
The inversion number equals the overtone label for the entries denoted by
$n=1,2,3$ in the tables, while the fundamental branch uses the uninverted
condition in Eq.~\eqref{eq:cf-fundamental}. The column headed $N$ in the
$C=1$ Leaver tables is the final recurrence depth used in the truncated right
continued fraction. For a candidate frequency we evaluated the normalized
continued-fraction residual after the root search and increased $N$ near the
quasi-resonant endpoint until changing $N$ did not affect the quoted conclusion
that the real part remains finite.

The same Frobenius framework can be combined with the midpoint
analytic-continuation prescription of Ref.~\cite{Rostworowski:2006bp} when
additional regular singularities obstruct the unit disk of convergence. In that
case one expands about the horizon, evaluates the local series at a midpoint
inside its convergence disk, re-expands there, and repeats the process until the
continued-fraction condition can be imposed at the final step. For the $C=1$ black-hole geometry used in the numerical scan below
this midpoint continuation is not needed, because the
asymptotic boundary factors above leave a single regular series on the interval
$0<x<1$.

\begin{table}[!ht]
\centering
\small
\begin{tabular}{cccc}
\toprule
\toprule
$\mu r_h$ & $\operatorname{Re}(\omega r_h)$ & $\operatorname{Im}(\omega r_h)$ & $N$ \\
\midrule
0.00 & 0.273386961360 & -0.410906550297 & 8000 \\
0.05 & 0.273434354064 & -0.409923564301 & 8000 \\
0.10 & 0.273559721182 & -0.406977848792 & 8000 \\
0.15 & 0.273712484378 & -0.402081385918 & 8000 \\
0.20 & 0.273808534633 & -0.395261899252 & 8000 \\
0.25 & 0.273733090776 & -0.386574966143 & 8000 \\
0.30 & 0.273350360832 & -0.376120431713 & 8000 \\
0.35 & 0.272525427018 & -0.364058528041 & 8000 \\
0.40 & 0.271160780593 & -0.350614624043 & 8000 \\
0.45 & 0.269238418230 & -0.336057891070 & 8000 \\
0.50 & 0.266843190924 & -0.320650138652 & 8000 \\
0.55 & 0.264145025434 & -0.304587182434 & 8000 \\
0.60 & 0.261347118596 & -0.287967990228 & 8000 \\
0.65 & 0.258633328502 & -0.270804551381 & 8000 \\
0.70 & 0.256140014075 & -0.253054746348 & 8000 \\
0.75 & 0.253952877900 & -0.234654677126 & 8000 \\
0.80 & 0.252116787036 & -0.215539629168 & 8000 \\
0.85 & 0.250648277130 & -0.195654062723 & 8000 \\
0.90 & 0.249546117688 & -0.174954563200 & 12000 \\
0.95 & 0.248799223852 & -0.153409515057 & 12000 \\
1.00 & 0.248382147225 & -0.130992666430 & 12000 \\
1.05 & 0.248296863709 & -0.107706053470 & 12000 \\
1.10 & 0.248502132273 & -0.083526923213 & 12000 \\
1.12 & 0.248663599549 & -0.073605890173 & 12000 \\
1.14 & 0.248869494007 & -0.063542562912 & 12000 \\
1.16 & 0.249119717854 & -0.053338524889 & 12000 \\
1.18 & 0.249401640441 & -0.042998978073 & 20000 \\
1.20 & 0.249737796305 & -0.032509532797 & 20000 \\
1.22 & 0.250096066188 & -0.021928265093 & 12000 \\
1.24 & 0.250443457440 & -0.011189267996 & 12000 \\
1.25 & 0.250810834304 & -0.005590737950 & 12000 \\
1.255 & 0.250819654928 & -0.002811038763 & 16000 \\
1.257 & 0.250780620025 & -0.001734725043 & 16000 \\
1.259 & 0.250767301846 & -0.000708775842 & 16000 \\
\bottomrule
\bottomrule
\end{tabular}
\caption{Fundamental massive scalar mode of the $C=1$ Bronnikov--Kim--2 black hole.}
\label{tab:bk2-c1-mass-scan}
\end{table}

\begin{table}[!ht]
\centering
\small
\begin{tabular}{cccc}
\toprule
\toprule
$\mu r_h$ & $\operatorname{Re}(\omega r_h)$ & $\operatorname{Im}(\omega r_h)$ & $N$ \\
\midrule
0.000 & 0.750846749 & -0.363872782 & 4000 \\
0.200 & 0.759893221 & -0.356502009 & 4000 \\
0.400 & 0.787375837 & -0.334185533 & 4000 \\
0.600 & 0.834436564 & -0.296257664 & 4000 \\
0.800 & 0.903457966 & -0.241389973 & 4000 \\
1.000 & 0.999378870 & -0.167037928 & 4000 \\
1.100 & 1.060668209 & -0.121045284 & 4000 \\
1.200 & 1.134829034 & -0.067523826 & 8000 \\
1.300 & 1.230343649 & -0.003287931 & 16000 \\
1.303 & 1.233736572 & -0.001102044 & 16000 \\
1.3034 & 1.234185962 & -0.000801255 & 16000 \\
\bottomrule
\bottomrule
\end{tabular}
\caption{Fundamental massive scalar mode with $\ell=1$ for the $C=1$
Bronnikov--Kim--2 black hole, computed from the Leaver continued fraction. The
near-endpoint rows were recomputed at several truncation depths; they establish
the persistence of finite real frequency as the damping becomes small.}
\label{tab:bk2-c1-ell-one-scan}
\end{table}

\begin{table}[!ht]
\centering
\small
\begin{tabular}{ccccc}
\toprule
\toprule
$n$ & $\mu r_h$ & $\operatorname{Re}(\omega r_h)$ & $\operatorname{Im}(\omega r_h)$ & $N$ \\
\midrule
1 & 0.00 & 0.180403632119 & -1.420565438441 & 4000 \\
1 & 0.50 & 0.171402618030 & -1.406254943189 & 6000 \\
1 & 1.00 & 0.152755557693 & -1.362988483421 & 4000 \\
1 & 1.50 & 0.137081547973 & -1.286220227287 & 6000 \\
1 & 2.00 & 0.128543377817 & -1.170840037803 & 4000 \\
1 & 2.50 & 0.126041470215 & -1.015781325443 & 6000 \\
1 & 3.00 & 0.127713294295 & -0.822718201148 & 4000 \\
1 & 3.50 & 0.132182268184 & -0.594226229738 & 6000 \\
1 & 4.00 & 0.138584828994 & -0.332851218883 & 6000 \\
1 & 4.40 & 0.144619305934 & -0.101598089242 & 6000 \\
1 & 4.50 & 0.146379745218 & -0.040932758487 & 10000 \\
1 & 4.54 & 0.147004797353 & -0.016382987834 & 10000 \\
1 & 4.56 & 0.147339832990 & -0.003861654469 & 10000 \\
1 & 4.565 & 0.147449330384 & -0.000860475471 & 24000 \\
\addlinespace
2 & 0.00 & 0.156463415534 & -2.439149806179 & 4000 \\
2 & 0.50 & 0.152094646582 & -2.432592194216 & 5000 \\
2 & 1.00 & 0.141218097563 & -2.411993569366 & 4000 \\
2 & 1.50 & 0.128507007985 & -2.374581080539 & 5000 \\
2 & 2.00 & 0.117637149924 & -2.316675045136 & 4000 \\
2 & 2.50 & 0.110026447961 & -2.235351388725 & 5000 \\
2 & 3.00 & 0.105584946581 & -2.129214707171 & 4000 \\
2 & 3.50 & 0.103693764963 & -1.998152459351 & 5000 \\
2 & 4.00 & 0.103709391799 & -1.842792582136 & 5000 \\
2 & 4.50 & 0.105134117997 & -1.664106250548 & 5000 \\
2 & 5.00 & 0.107586624238 & -1.463170301646 & 6000 \\
2 & 5.50 & 0.110839098394 & -1.241033788259 & 6000 \\
2 & 6.00 & 0.114706158477 & -0.998681118415 & 6000 \\
2 & 6.50 & 0.119055031480 & -0.737029474468 & 6000 \\
2 & 7.00 & 0.123799959402 & -0.456920957425 & 6000 \\
2 & 7.50 & 0.128851829071 & -0.159219958171 & 8000 \\
2 & 7.70 & 0.130918663337 & -0.035256073127 & 8000 \\
2 & 7.75 & 0.131467253466 & -0.003879898101 & 12000 \\
2 & 7.755 & 0.131540423484 & -0.000689410155 & 24000 \\
\bottomrule
\bottomrule
\end{tabular}
\caption{First two overtones of the $C=1$ Bronnikov--Kim--2  spectrum computed by the Leaver method.}
\label{tab:bk2-c1-overtone-scan}
\end{table}

\begin{table}[!ht]
\centering
\small
\begin{tabular}{cccc}
\toprule
\toprule
$\mu r_h$ & $\operatorname{Re}(\omega r_h)$ & $\operatorname{Im}(\omega r_h)$ & $N$ \\
\midrule
0.00 & 0.145855367505 & -3.449950095763 & 12000 \\
0.50 & 0.143162027885 & -3.445900326400 & 5000 \\
1.00 & 0.135964999332 & -3.433208817078 & 5000 \\
1.50 & 0.126385656409 & -3.410310737406 & 5000 \\
2.00 & 0.116661250725 & -3.374944371533 & 5000 \\
2.50 & 0.108302599061 & -3.324773653165 & 5000 \\
3.00 & 0.101932191954 & -3.257963400546 & 5000 \\
3.50 & 0.097558698369 & -3.173415021022 & 5000 \\
4.00 & 0.094910750817 & -3.070696186585 & 5000 \\
4.50 & 0.093646472192 & -2.949851568350 & 5000 \\
5.00 & 0.093463773494 & -2.811208977105 & 5000 \\
5.50 & 0.094118444914 & -2.655263233066 & 5000 \\
6.00 & 0.095413576613 & -2.482565887564 & 5000 \\
6.50 & 0.097228376448 & -2.293678303163 & 7000 \\
7.00 & 0.099441579581 & -2.089186788692 & 7000 \\
7.50 & 0.101994467157 & -1.869602636372 & 7000 \\
8.00 & 0.104835470690 & -1.635484397057 & 7000 \\
8.50 & 0.107878400039 & -1.387260833755 & 7000 \\
9.00 & 0.111181956928 & -1.125423914658 & 7000 \\
9.50 & 0.114581294981 & -0.850374460210 & 7000 \\
10.00 & 0.118235003509 & -0.562526890728 & 9000 \\
10.50 & 0.121991033162 & -0.262172518156 & 9000 \\
10.80 & 0.124277960361 & -0.076135633416 & 12000 \\
10.90 & 0.125024545065 & -0.013244713354 & 12000 \\
10.91 & 0.125143787215 & -0.006855751032 & 16000 \\
10.92 & 0.125185641755 & -0.000531190649 & 16000 \\
\bottomrule
\bottomrule
\end{tabular}
\caption{Third overtone of the $C=1$ Bronnikov--Kim--2 spectrum computed by the Leaver method.}
\label{tab:bk2-c1-third-overtone-scan}
\end{table}

\section{Numerical spectra}
\label{sec:numerical}

The numerical analysis has two complementary parts. On the black-hole side we
use continued fractions to push the $C=1$ Bronnikov--Kim--2 black-hole spectrum
close to quasi-resonant endpoints. On the wormhole side we use a finite
Chebyshev pseudospectral problem based on a
smooth one-exterior symmetry reduction to test whether the same massive-field
mechanism produces long-lived ringing in two-ended brane-world geometries.
The normalization of the numerical data is specified here, where
the tables are introduced. For black-hole tables we quote the dimensionless
combinations $\omega r_h$ and $\mu r_h$, with $r_h$ the event-horizon radius;
this is equivalent to setting $r_h=1$ in the numerical scan. For wormhole
tables there is no horizon radius, so the length scale is fixed by $2M=1$ in
BK--1 and by setting the scale $a$ in Eq.~\eqref{eq:bk2} to unity in BK--2.

\begin{table}[tp]
\centering
\scriptsize
\begin{tabular}{ccccc}
\toprule
\toprule
$C$ & $\omega_{\rm H}r_h$ & $\omega_{\rm TD}r_h$ & $|\Delta\omega|r_h$ & $\delta_{\rm rel}$ \\
\midrule
1.50 & $0.536920-0.342922i$ & $0.53695-0.34301i$ & $9.3\times10^{-5}$ & $0.0145\%$ \\
1.00 & $0.576670-0.317494i$ & $0.57668-0.31754i$ & $4.7\times10^{-5}$ & $0.0072\%$ \\
0.50 & $0.617255-0.275489i$ & $0.61726-0.27547i$ & $2.0\times10^{-5}$ & $0.0029\%$ \\
0.10 & $0.641578-0.224046i$ & $0.64156-0.22399i$ & $5.9\times10^{-5}$ & $0.0087\%$ \\
\bottomrule
\bottomrule
\end{tabular}
\caption{Electromagnetic $\ell=1$ black-hole-side modes of the Bronnikov--Kim--2 geometry. The Chebyshev--Hill values $\omega_{\rm H}$ are compared with the time-domain values $\omega_{\rm TD}$ quoted in Table~III of Ref.~\cite{Bronnikov:2019sbx}. Here $\Delta\omega=\omega_{\rm H}-\omega_{\rm TD}$ and $\delta_{\rm rel}=100|\Delta\omega|/|\omega_{\rm TD}|$. The Chebyshev--Hill calculation used $N_{\rm H}=60$ and endpoint clipping $\varepsilon=10^{-3}$.}
\label{tab:em-hill-check}
\end{table}

\begin{table}[tp]
\centering
\scriptsize
\begin{tabular}{cccc}
\toprule
\toprule
$\mu r_h$ & $\operatorname{Re}(\omega r_h)$ & $-\operatorname{Im}(\omega r_h)$ & $\Delta_N r_h$ \\
\midrule
0.00 & 0.269927 & 0.331205 & $6.1\times10^{-6}$ \\
0.01 & 0.269944 & 0.331162 & $6.1\times10^{-6}$ \\
0.02 & 0.269995 & 0.331033 & $6.1\times10^{-6}$ \\
0.03 & 0.270079 & 0.330818 & $6.2\times10^{-6}$ \\
0.04 & 0.270197 & 0.330516 & $6.3\times10^{-6}$ \\
0.05 & 0.270347 & 0.330129 & $6.5\times10^{-6}$ \\
0.10 & 0.271591 & 0.326901 & $7.9\times10^{-6}$ \\
0.15 & 0.273603 & 0.321520 & $1.1\times10^{-5}$ \\
0.20 & 0.276290 & 0.313992 & $1.7\times10^{-5}$ \\
0.25 & 0.279519 & 0.304345 & $3.0\times10^{-5}$ \\
0.30 & 0.283123 & 0.292647 & $5.9\times10^{-5}$ \\
0.35 & 0.286933 & 0.279025 & $1.3\times10^{-4}$ \\
0.40 & 0.290815 & 0.263619 & $2.8\times10^{-4}$ \\
0.45 & 0.294533 & 0.246435 & $6.2\times10^{-4}$ \\
0.50 & 0.297027 & 0.227752 & $1.4\times10^{-3}$ \\
0.55 & 0.297466 & 0.211331 & $3.4\times10^{-3}$ \\
\bottomrule
\bottomrule
\end{tabular}
\caption{Massive scalar $\ell=0$ fundamental branch for the Bronnikov--Kim--2 black hole with $C=1/2$, computed with the Chebyshev--Hill pseudospectral truncation. The increasing value of $\Delta_N$ in the last rows indicates stronger finite-matrix sensitivity; points beyond the range shown were not used.}
\label{tab:c-half-massive-ps}
\end{table}

\begin{table}[tp]
\centering
\textbf{(a) $C=3/2$}\par\vspace{0.4ex}
\begin{tabular}{@{}cccc@{}}
\toprule
\toprule
$\mu r_h$ & $\operatorname{Re}\omega r_h$ & $-\operatorname{Im}\omega r_h$ & $\Delta_N r_h$ \\
\midrule
0.00 & 0.506853 & 0.294554 & $1.1\times10^{-9}$ \\
0.10 & 0.508730 & 0.289625 & $1.1\times10^{-9}$ \\
0.20 & 0.514576 & 0.274285 & $1.2\times10^{-9}$ \\
0.30 & 0.525388 & 0.246541 & $1.3\times10^{-9}$ \\
0.40 & 0.545280 & 0.201882 & $1.6\times10^{-9}$ \\
0.50 & 0.591395 & 0.139162 & $2.1\times10^{-9}$ \\
0.60 & 0.671996 & 0.088647 & $1.7\times10^{-9}$ \\
0.70 & 0.760766 & 0.058540 & $1.2\times10^{-9}$ \\
0.80 & 0.850121 & 0.039048 & $9.3\times10^{-10}$ \\
0.90 & 0.939665 & 0.025556 & $6.5\times10^{-10}$ \\
1.00 & 1.029705 & 0.015950 & $5.2\times10^{-10}$ \\
1.10 & 1.120568 & 0.009094 & $3.8\times10^{-10}$ \\
1.20 & 1.212548 & 0.004314 & $2.4\times10^{-10}$ \\
1.24 & 1.249717 & 0.002885 & $2.0\times10^{-10}$ \\
1.28 & 1.287133 & 0.001702 & $1.5\times10^{-10}$ \\
1.30 & 1.305940 & 0.001200 & $1.3\times10^{-10}$ \\
1.32 & 1.324818 & 0.000754 & $1.3\times10^{-10}$ \\
1.34 & 1.343769 & 0.000366 & $1.3\times10^{-10}$ \\
1.35 & 1.353274 & 0.000193 & $1.4\times10^{-10}$ \\
1.36 & 1.362798 & 0.0000344 & $1.8\times10^{-10}$ \\
1.361 & 1.363752 & 0.0000193 & $2.0\times10^{-10}$ \\
1.362 & 1.364706 & 0.00000441 & $2.2\times10^{-10}$ \\
\bottomrule
\bottomrule
\end{tabular}
\par\smallskip
\textbf{(b) $C=2$}\par\vspace{0.4ex}
\begin{tabular}{@{}cccc@{}}
\toprule
\toprule
$\mu r_h$ & $\operatorname{Re}\omega r_h$ & $-\operatorname{Im}\omega r_h$ & $\Delta_N r_h$ \\
\midrule
0.00 & 0.562888 & 0.308846 & $1.3\times10^{-9}$ \\
0.10 & 0.564730 & 0.304318 & $1.3\times10^{-9}$ \\
0.20 & 0.570474 & 0.290314 & $1.4\times10^{-9}$ \\
0.30 & 0.581046 & 0.265414 & $1.6\times10^{-9}$ \\
0.40 & 0.599499 & 0.226817 & $2.2\times10^{-9}$ \\
0.50 & 0.635689 & 0.172928 & $2.3\times10^{-9}$ \\
0.60 & 0.701333 & 0.118812 & $2.1\times10^{-9}$ \\
0.70 & 0.783389 & 0.081020 & $1.8\times10^{-9}$ \\
0.80 & 0.869451 & 0.055754 & $1.3\times10^{-9}$ \\
0.90 & 0.956659 & 0.037995 & $1.1\times10^{-9}$ \\
1.00 & 1.044606 & 0.025065 & $8.9\times10^{-10}$ \\
1.10 & 1.133360 & 0.015496 & $6.2\times10^{-10}$ \\
1.20 & 1.223090 & 0.008414 & $4.5\times10^{-10}$ \\
1.25 & 1.268384 & 0.005622 & $3.7\times10^{-10}$ \\
1.30 & 1.314000 & 0.003259 & $3.2\times10^{-10}$ \\
1.34 & 1.350747 & 0.001647 & $2.7\times10^{-10}$ \\
1.36 & 1.369211 & 0.000927 & $2.5\times10^{-10}$ \\
1.37 & 1.378468 & 0.000586 & $2.5\times10^{-10}$ \\
1.38 & 1.387742 & 0.000259 & $2.2\times10^{-10}$ \\
1.386 & 1.393314 & 0.0000685 & $2.2\times10^{-10}$ \\
1.388 & 1.395173 & 0.00000597 & $2.0\times10^{-10}$ \\
\bottomrule
\bottomrule
\end{tabular}
\caption{Massive scalar $\ell=0$ oscillatory branches for the Bronnikov--Kim--2
black hole with $C=3/2$ and $C=2$, computed with the same
Chebyshev--Hill pseudospectral truncation as Table~\ref{tab:c-half-massive-ps}.
The values use $N_{\rm H}=60$, and
$\Delta_N=|\omega_{N_{\rm H}=60}-\omega_{N_{\rm H}=50}|$. For $C=3/2$ the
branch crosses the real axis between $\mu r_h=1.362$ and $1.363$; for $C=2$ it
crosses between $\mu r_h=1.388$ and $1.3885$.}
\label{tab:positive-c-massive-ps}
\end{table}

\begin{table}
\centering
\scriptsize
\begin{tabular}{ccccc}
\toprule
\toprule
Geometry & $\mu$ & $\operatorname{Re}\omega$ & $-\operatorname{Im}\omega$ & $\Delta_{\rm ps}$ \\
\midrule
BK--1, $r_0=1.3$ & 0.00 & 0.139890080 & 0.012163709 & $2.8\times10^{-8}$ \\
BK--1, $r_0=1.3$ & 0.03 & 0.140091521 & 0.011944814 & $4.9\times10^{-6}$ \\
BK--1, $r_0=1.3$ & 0.06 & 0.140696994 & 0.011289918 & $2.5\times10^{-5}$ \\
BK--1, $r_0=1.3$ & 0.09 & 0.141606958 & 0.010186092 & $6.6\times10^{-5}$ \\
BK--1, $r_0=1.3$ & 0.12 & 0.142897611 & 0.008670160 & $4.1\times10^{-4}$ \\
\addlinespace
BK--2, $C=-0.3$ & 0.00 & 0.331974693 & 0.073155916 & $2.5\times10^{-11}$ \\
BK--2, $C=-0.3$ & 0.05 & 0.332928476 & 0.072323629 & $2.1\times10^{-11}$ \\
BK--2, $C=-0.3$ & 0.10 & 0.335791094 & 0.069834466 & $4.2\times10^{-11}$ \\
BK--2, $C=-0.3$ & 0.15 & 0.340564304 & 0.065706391 & $1.6\times10^{-11}$ \\
BK--2, $C=-0.3$ & 0.20 & 0.347243932 & 0.059949474 & $3.1\times10^{-11}$ \\
BK--2, $C=-0.3$ & 0.25 & 0.355806330 & 0.052522618 & $5.2\times10^{-11}$ \\
BK--2, $C=-0.3$ & 0.30 & 0.366169421 & 0.043220306 & $2.7\times10^{-10}$ \\
BK--2, $C=-0.3$ & 0.35 & 0.377968509 & 0.031256785 & $5.4\times10^{-9}$ \\
BK--2, $C=-0.3$ & 0.38 & 0.384604859 & 0.021564515 & $5.1\times10^{-7}$ \\
BK--2, $C=-0.3$ & 0.39 & 0.385732368 & 0.017779836 & $1.2\times10^{-5}$ \\
BK--2, $C=-0.3$ & 0.40 & 0.386015608 & 0.014895110 & $2.8\times10^{-4}$ \\
\bottomrule
\bottomrule
\end{tabular}
\caption{Even-symmetry-sector fundamental
massive scalar modes of representative Bronnikov--Kim wormholes. The
geometry column gives the parameter choice for each family; for BK--1 the
normalization is $2M=1$. The quoted sensitivity $\Delta_{\rm ps}$ is the
largest change of the complex frequency in the compactified Chebyshev
calculations used for each block.}
\label{tab:bk-wormhole-massive-ps}
\end{table}

\subsection{Chebyshev--Hill check away from $C=1$}

Before following massive branches, it is useful to check the rationalized
positive-$C$ equation away from the $C=1$ black-hole case.
For this purpose we used a finite Chebyshev--Hill truncation of the regular
equation for $u$. After the asymptotic
boundary factor has been divided out, the equation has the form
\begin{equation}
 p_2(x,\omega)u''+p_1(x,\omega)u'+p_0(x,\omega)u=0,
 \label{eq:hill-regular-equation}
\end{equation}
where the $p_j$ are rational functions of the compact coordinate and polynomial
functions of $\omega$ for the massless problem. In practice we evaluate
Eq.~\eqref{eq:hill-regular-equation} on the Chebyshev--Lobatto points
\begin{equation}
 x_j=\frac12\left[1-\cos\left(\frac{\pi j}{N_{\rm H}}\right)\right],
 \qquad j=0,\ldots,N_{\rm H},
\end{equation}
or, in the explicit matrix implementation below, on the slightly clipped interval
$\varepsilon\le x\le1-\varepsilon$ to avoid evaluating the singular endpoint behaviour that has already been
absorbed into $F$ exactly. If $D$ is the Chebyshev differentiation matrix, the
truncated equation is
\begin{equation}
 \left[P_2(\omega)D^2+P_1(\omega)D+P_0(\omega)\right]{\bf u}=0,
 \label{eq:hill-matrix}
\end{equation}
where $P_j$ are diagonal matrices with entries $p_j(x_i,\omega)$. For massless
fields the outgoing wave number is $k=\omega$, so the matrix in
Eq.~\eqref{eq:hill-matrix} is quadratic in $\omega$,
\begin{equation}
 M(\omega)=M_0+\omega M_1+\omega^2 M_2.
\end{equation}
The finite Chebyshev--Hill spectrum is therefore obtained from the linearized
generalized eigenvalue problem
\begin{equation}
 \begin{pmatrix}
  0&I\\ -M_0&-M_1
 \end{pmatrix}
 \begin{pmatrix}{\bf u}\\ \omega {\bf u}\end{pmatrix}
 =\omega
 \begin{pmatrix}
  I&0\\ 0&M_2
 \end{pmatrix}
 \begin{pmatrix}{\bf u}\\ \omega {\bf u}\end{pmatrix}.
 \label{eq:hill-linearized}
\end{equation}
This is not as efficient as the continued fraction for high-precision
quasi-resonant endpoints, but it is a convenient independent check of the
rationalized positive-$C$ equation.

As a check we reproduced the electromagnetic $\ell=1$ black-hole-side modes
listed in Table~III of Ref.~\cite{Bronnikov:2019sbx}. For electromagnetic
perturbations the master potential reduces to
\begin{equation}
 V_{\rm em}(r)=A(r)\frac{\ell(\ell+1)}{r^2}.
 \label{eq:em-potential}
\end{equation}
Using Eq.~\eqref{eq:bk2-x-arbitrary-c} and the massless version of
Eq.~\eqref{eq:general-c-x-leaver-factor}, the Chebyshev--Hill roots shown in
Table~\ref{tab:em-hill-check} reproduce the time-domain black-hole values of
Ref.~\cite{Bronnikov:2019sbx}. The $C=0$ row of Table~III of that reference is
not included here because the horizon is then degenerate and the non-extremal
factor $x^{-i\omega/(2\sqrt{C})}$ is no longer applicable.

This comparison tests the same rationalized positive-$C$ radial equation against
previously published data before any massive-field continuation is attempted. The largest absolute difference in
Table~\ref{tab:em-hill-check} is below $10^{-4}/r_h$, and the largest
relative difference is about $1.5\times10^{-2}\%$. These differences are at the
level expected from comparing a finite spectral truncation with frequencies read
from time-domain profiles. The table also shows why the Chebyshev--Hill calculation is not used for the
high-precision endpoint data: the finite matrix contains additional
discretization roots, so the physical mode must be identified by stability under
increasing $N_{\rm H}$ and decreasing $\varepsilon$. The continued fraction
remains the more reliable tool for the high-order tail treatment used for the
near-resonant $C=1$ massive data below.

\subsection{Massive pseudospectral calculation at $C=1/2$}

The same pseudospectral implementation can be used for the massive scalar
equation. The massive wave number $k=\sqrt{\omega^2-\mu^2}$ is included in
the asymptotic boundary factor, and the
collocation equations are solved by continuation in $\mu$, with the vector of
values of $u$ normalized at an interior Chebyshev point. Table~\ref{tab:c-half-massive-ps}
shows the fundamental $\ell=0$ branch for the positive-$C$ black-hole geometry
with $C=1/2$. The quoted values use $N_{\rm H}=60$ interior Chebyshev points,
and the last column estimates the truncation sensitivity from
\begin{equation}
 \Delta_N=\left|\omega_{N_{\rm H}=60}-\omega_{N_{\rm H}=50}\right|.
\end{equation}
In the collocation solve the unknowns are the complex frequency and the values
of the regular function $u$ on the clipped Chebyshev--Lobatto grid.
The asymptotic boundary factors in
Eq.~\eqref{eq:general-c-x-leaver-factor} are divided out before the matrix is
formed, and one interior component of $u$ is fixed to unity
to remove the homogeneous normalization freedom. The same continuation in
$\mu$ and the same definition of $\Delta_N$ are used for the positive-$C$
finite-matrix data in Table~\ref{tab:positive-c-massive-ps}.

\begin{figure}[tp]
\centering
\includegraphics[width=0.95\columnwidth]{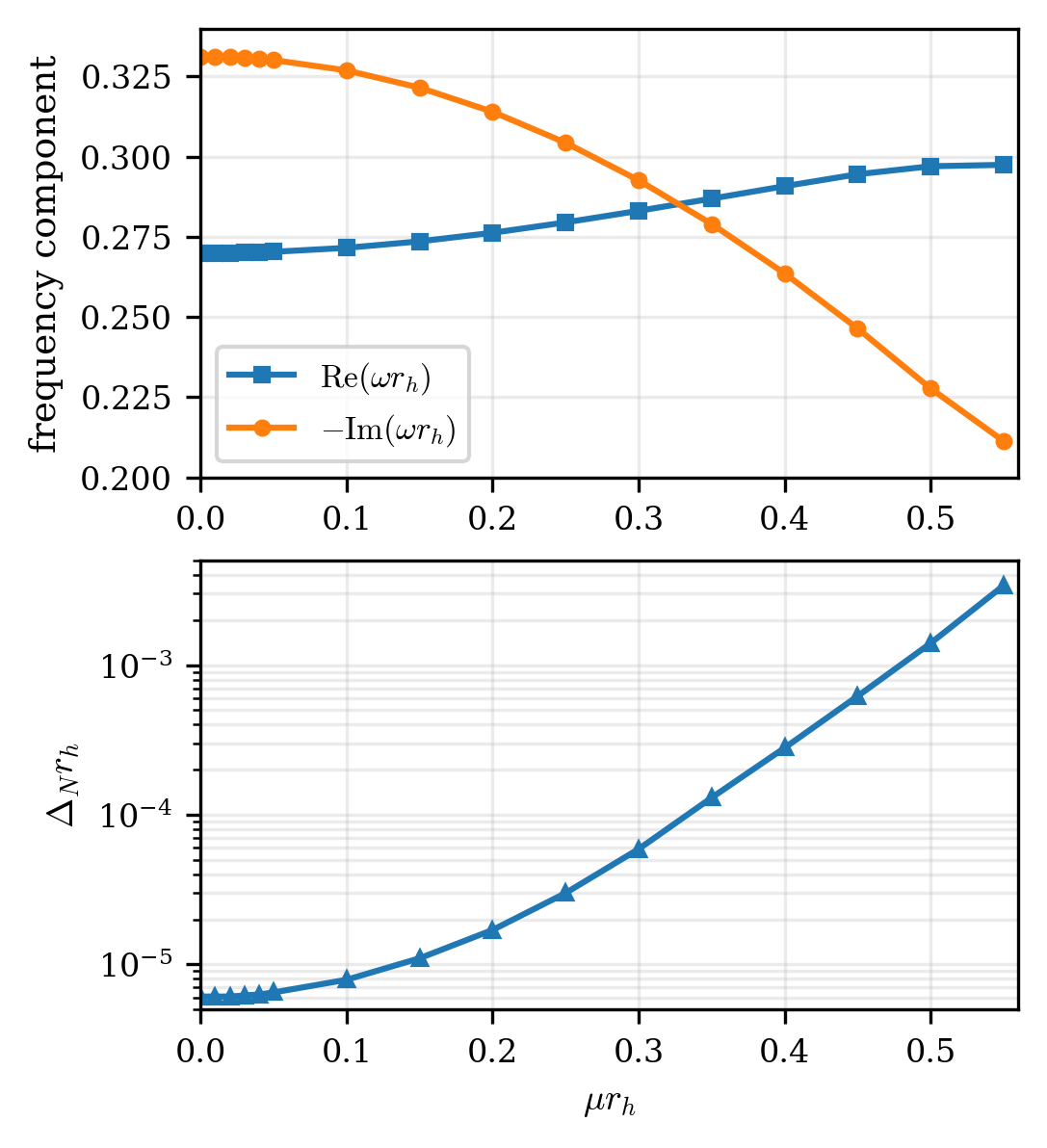}
\caption{Graphical version of the $C=1/2$ data in
Table~\ref{tab:c-half-massive-ps}. The upper panel shows the real part and
damping rate of the fundamental scalar mode, while the lower panel shows the
truncation sensitivity estimate on a logarithmic scale.}
\label{fig:c-half-massive-ps}
\end{figure}

The $C=1/2$ scan, shown graphically in
Fig.~\ref{fig:c-half-massive-ps}, shows the same qualitative trend as the
$C=1$ black-hole data:
increasing the field mass lowers the damping rate, while the real part remains
finite throughout the stable part of the calculation. This finite-matrix
calculation confirms that the rationalized positive-$C$
equation is already useful away from the
$C=1$ black-hole case. For $\mu r_h\gtrsim0.55$ the finite collocation
matrix develops competing small singular values and stronger dependence on $N_{\rm H}$, so a Leaver or midpoint
continuation treatment would be preferable for locating a quasi-resonant
endpoint at this value of $C$.

We also repeated the same finite-matrix continuation for the
oscillatory scalar branch at $C=3/2$ and $C=2$. In these cases the additional
zero of $B_2$ lies beyond infinity in the compactified coordinate, and the
branch is considerably easier to follow. Table~\ref{tab:positive-c-massive-ps}
shows that the damping rate can be driven below
$10^{-5}/r_h$ before the finite truncation crosses the real axis. For $C=3/2$
the last stable point found in this scan is $\mu r_h=1.362$, where
$\omega r_h\simeq1.36471-4.41\times10^{-6}i$.

For $C=2$ the last stable point found in this scan is $\mu r_h=1.388$, where
$\omega r_h\simeq1.39517-5.97\times10^{-6}i$. These positive-$C$ tests
therefore support the same qualitative conclusion as the $C=1$ continued-fraction
calculation: increasing the field mass suppresses the damping while leaving a
finite oscillation frequency on the tracked branch.

\subsection{Wormhole-side long-lived modes}

The same strategy of factoring out
the known outer-boundary wave behaviour can also be used on the wormhole side, where the physical problem is a smooth two-ended scattering problem.
The throat is a regular interior point. In the time-domain formulation the
scalar field evolves smoothly through it, and in the frequency-domain problem no
reflecting or absorbing throat boundary is imposed. For the reflection-symmetric
examples tabulated here, we use only the even sector as a computational
half-domain reduction of the full two-ended problem: the one-exterior solution is
smoothly continued through the throat to the second exterior. This half-domain
matching condition is not a wall and does not make the wormhole nontraversable.
The far asymptotic end is treated with the outgoing massive factor
$\exp(ikr_*)$, $k=\sqrt{\omega^2-\mu^2}$. The remaining regular function is
collocated on a compactified Chebyshev grid. On the exterior side we compactify
the half-line by a smooth map of the form $\rho=L\xi/(1-\xi)$,
$0\le\xi<1$, factor out $\exp(ikr_*)$ at $\xi=1$, and compare nearby choices of
the map scale $L$, the endpoint clipping, and the collocation order to estimate
$\Delta_{\rm ps}$.
To avoid the nearly continuum-like box roots that can appear in a finite
truncated interval, Table~\ref{tab:bk-wormhole-massive-ps} follows the nonzero,
stable even-symmetry-sector fundamental root under changes of the
compactification map, endpoint clipping, and number of points.

Quasinormal modes of massive scalar fields around traversable wormholes
were also considered in Ref.~\cite{Churilova:2019qph}, which therefore provides
a useful point of comparison for the Bronnikov--Kim wormhole side. That work
showed that massive fields on wormhole backgrounds can support
arbitrarily long-lived QNMs. For the present discussion, the relevant lesson is
dynamical. As Fig.~\ref{fig:bk-effective-potentials} shows, the
Bronnikov--Kim wormhole examples already have a two-sided barrier structure; the
important additional feature emphasized in Ref.~\cite{Churilova:2019qph} is
that, in the time-domain signal, the massive-field tail can dominate before the
long-lived ringing stage is clearly seen. Thus the usual ordering ``ringdown
first, tail later'' may be reversed. Our calculation is frequency-domain mode
tracking and shows that the damping of the
even-symmetry branch decreases as the scalar mass is raised. A separate time-domain evolution or excitation analysis
is therefore needed to identify the interval over which this branch gives a
clean Bronnikov--Kim wormhole ringdown signal.

For illustration we chose one wormhole in each Bronnikov--Kim family: the BK--1
wormhole with $r_0=1.3$, and the BK--2 wormhole with $C=-0.3$ in
Eq.~\eqref{eq:bk2}. Frequencies are quoted in the same units as the black-hole
tables. In BK--1 these units are fixed by $2M=1$, while in BK--2 they are fixed
by the scale already used in Eq.~\eqref{eq:bk2}. The BK--2 metric extends the
mass scan to $\mu=0.40$, where the damping rate is about five times smaller than
at $\mu=0$. Larger masses were not used because the finite matrix begins to
show stronger sensitivity to the compactification and nearby small singular
values.

Table~\ref{tab:bk-wormhole-massive-ps} shows that the wormhole-side modes follow
the same qualitative massive-field trend found on the black-hole side:
increasing the field mass lowers the damping rate while the oscillation
frequency remains finite. For the extended BK--2 branch the damping decreases
from $7.32\times10^{-2}$ to $1.49\times10^{-2}$, whereas
$\operatorname{Re}\omega$ increases from about $0.332$ to $0.386$. These
numbers should be read as a pseudospectral mode-tracking calculation rather than
a high-precision survey of the full two-ended problem.
The complementary symmetry sector and the
complete set of trapped echo-like wormhole resonances require a separate
analysis.

\subsection{Massive spectrum for $C=1$}

We now specialize to scalar perturbations of the $C=1$ black-hole geometry.
Frequencies are quoted in units of the horizon radius. This black-hole branch
is the most direct comparison with the massive scalar calculation of
Ref.~\cite{Lutfuoglu:2026cfm}, and it is also the case in which the Leaver
method is most efficient: the asymptotic
boundary factors are known analytically and the nearly resonant tail can be
stabilized by increasing the recurrence depth. The data below therefore include
both the monopole sequence and the dipole fundamental branch. Table~\ref{tab:bk2-c1-mass-scan} gives the $\ell=0$
fundamental mode, Tables~\ref{tab:bk2-c1-overtone-scan} and
\ref{tab:bk2-c1-third-overtone-scan} give the first three $\ell=0$ overtones,
and Table~\ref{tab:bk2-c1-ell-one-scan} gives the $\ell=1$ fundamental mode.
The overtones were also verified in the uninverted continued fraction, which
removes spurious roots introduced by inversion denominators.

The fundamental branch already shows the main point. At $\mu r_h=1.259$ the
damping rate is below $10^{-3}$, while the real frequency remains close to
$0.25/r_h$. The real part varies only mildly over the whole scan: it first
increases slightly, then decreases, and finally turns upward as the damping rate
collapses. Thus the branch becomes long-lived without approaching
$\operatorname{Re}\omega=0$.

The dipole branch behaves in the same way as the monopole branches. The
centrifugal barrier raises the oscillation frequency, but it does not change the
massive-field trend: the damping decreases by more than two orders of magnitude,
and at the last stable point shown in Table~\ref{tab:bk2-c1-ell-one-scan} the
mode has $\omega r_h\simeq1.23419-8.0\times10^{-4}i$, with no tendency toward
$\operatorname{Re}\omega=0$.

\begin{figure}[tp]
\centering
\includegraphics[width=0.95\columnwidth]{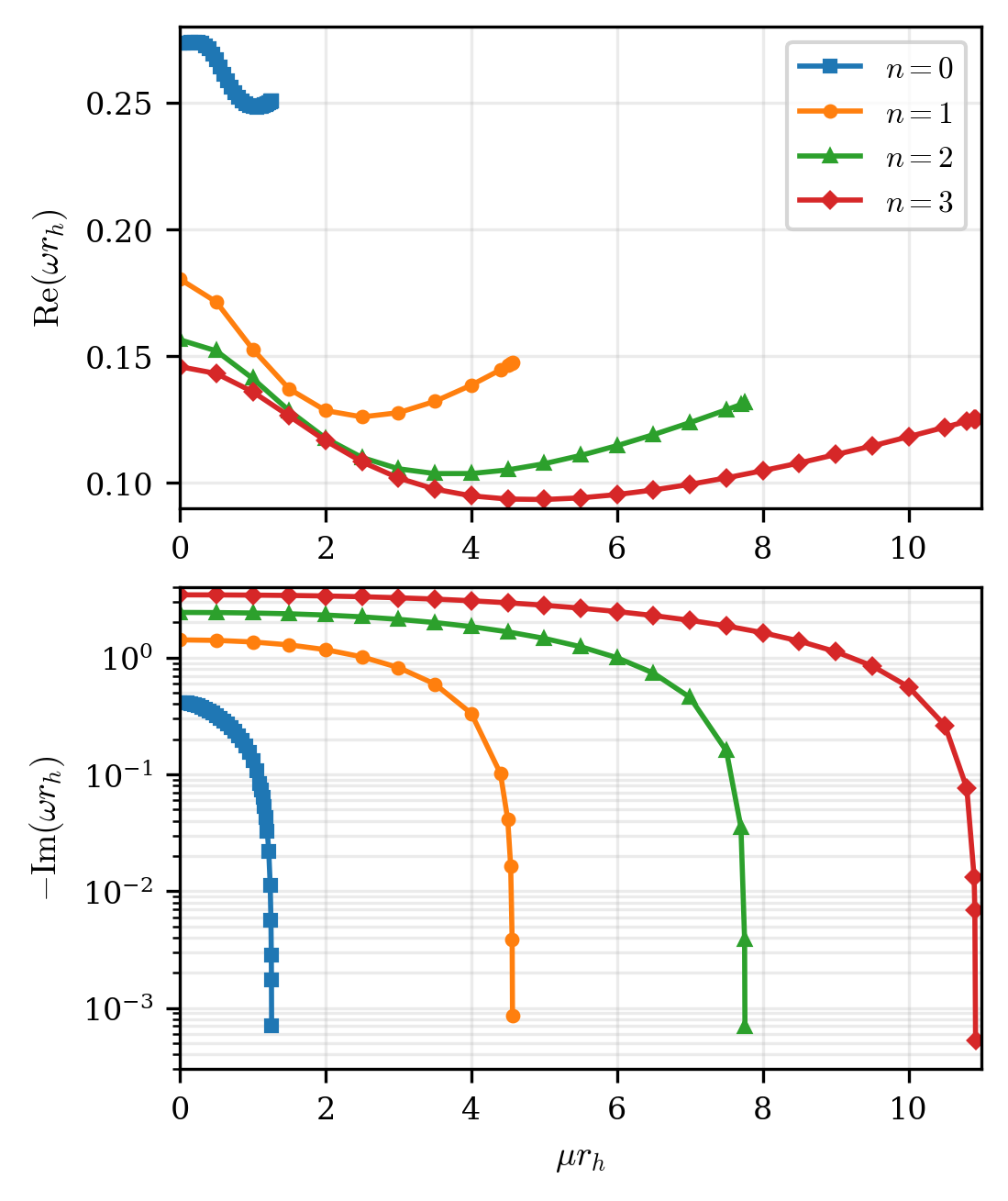}
\caption{Mass dependence of the fundamental mode and first three overtones for
the $C=1$ Bronnikov--Kim--2 scalar problem. The damping-rate panel uses a
logarithmic scale to show the approach to quasi-resonant endpoints.}
\label{fig:bk2-c1-overtone-scan}
\end{figure}

Fig.~\ref{fig:bk2-c1-overtone-scan} summarizes the four branches together. The
first overtone reaches $\omega r_h\simeq0.14745-0.00086i$ at
$\mu r_h=4.565$; the second reaches $\omega r_h\simeq0.13154-0.00069i$ at
$\mu r_h=7.755$; and the third reaches
$\omega r_h\simeq0.12519-0.00053i$ at $\mu r_h=10.92$. These endpoints,
like the fundamental endpoint, have small damping and nonzero real frequency.
The endpoint mass increases rapidly with overtone number, whereas the limiting
real part decreases more slowly and remains of order $10^{-1}/r_h$ for all
branches shown. The replacement of the fundamental branch by overtones
therefore does not lead to the $\operatorname{Re}\omega\to0$ behaviour reported
for the CFM geometry.

The qualitative distinction can be traced to the effective radial problem rather
than to the brane-world label alone. For this Bronnikov--Kim black hole, the mass term raises the asymptotic value of the potential and permits a sequence of
long-lived outgoing solutions. The numerical data show no indication that the
oscillatory part of these solutions is driven to zero before the damping rate
vanishes. Consequently, the CFM zero-real-frequency behaviour should be viewed
as a special feature of that geometry or of a particular spectral branch, not as
a universal consequence of adding a brane correction to a spherical black hole.

The wormhole calculation broadens this conclusion. It shows that long-lived
massive-field ringing is not tied to an event horizon: regular Bronnikov--Kim
throats also support modes whose damping decreases as the field mass is raised.
The present finite-matrix data identify the same physical tendency on the wormhole side of the
brane-world parameter space.

\subsection{Eikonal WKB limit}
\label{subsec:eikonal-wkb}

The high-multipole limit gives a simple analytic check on the black-hole side of
the calculation. Here ``high multipole'' means $\ell\gg1$, where $\ell$ is the
spherical-harmonic index introduced in the separation of variables. In this
limit the angular part of the scalar potential dominates the barrier, and the
potential has a single maximum outside the horizon for both Bronnikov--Kim
black-hole metrics considered here. We therefore apply the first-order
Wentzel--Kramers--Brillouin (WKB) condition at the top of the barrier,
\begin{equation}
 \omega^2=V_0-i\left(n+\frac12\right)
 \sqrt{-2V_{0,r_*r_*}},
 \label{eq:first-order-wkb-eikonal}
\end{equation}
where $\omega$ is the complex QNM frequency, $n=0,1,2,\ldots$ is the overtone
number, $i$ is the imaginary unit, $V_0=V(r_p)$ is the value of the effective
potential $V(r)$ at its peak, $r_p$ is the peak radius, and
$V_{0,r_*r_*}=d^2V/dr_*^2|_{r_p}$ is the second derivative with respect to the
tortoise coordinate $r_*$ at that peak. The coordinate $r$ below is the areal
radius, while $A(r)$ and $B(r)$ are the metric functions in
Eq.~\eqref{eq:schrodinger}. It is useful to define
\begin{equation}
 L=\ell+\frac12,
 \qquad
 H(r)=\frac{A(r)}{r^2}.
\end{equation}
Here $L$ is the eikonal angular parameter and $H(r)$ is the leading angular part
of the effective potential divided by $L^2$. For fixed scalar-field mass $\mu$
and $L\gg1$, Eq.~\eqref{eq:general-potential} gives
$V(r)=L^2H(r)+\mathcal{O}(L^0)$, where $\mathcal{O}(L^0)$ denotes terms that remain finite as
$L\to\infty$. The leading peak location is therefore determined by
\begin{equation}
 H'(r_c)=0,
 \label{eq:eikonal-peak-condition}
\end{equation}
where the prime denotes $d/dr$ and $r_c$ is the leading eikonal peak radius,
equivalently the radius of the unstable null circular orbit. At this radius,
\begin{equation}
 V_{0,r_*r_*}=L^2A(r_c)B(r_c)H''(r_c)+\mathcal{O}(L^0),
\end{equation}
where $H''=d^2H/dr^2$. Expanding
Eq.~\eqref{eq:first-order-wkb-eikonal} in powers of $L$ gives the standard
eikonal form
\begin{equation}
 \omega_{\ell n}=L\Omega_c-i\left(n+\frac12\right)\lambda_c+\mathcal{O}(L^{-1}),
 \label{eq:eikonal-master-formula}
\end{equation}
where $\omega_{\ell n}$ is the QNM frequency with multipole number $\ell$ and
overtone number $n$, while $O(L^{-1})$ denotes terms suppressed by one power of
$L$. The quantities $\Omega_c$ and $\lambda_c$ are
\begin{align}
 \Omega_c&=\sqrt{H(r_c)}=\frac{\sqrt{A(r_c)}}{r_c},\notag\\
 \lambda_c&=\sqrt{-\frac{A(r_c)B(r_c)H''(r_c)}{2H(r_c)}}.
 \label{eq:eikonal-omega-lambda}
\end{align}
Here $\Omega_c$ is the coordinate angular frequency of the unstable null circular
orbit, and $\lambda_c$ is its Lyapunov exponent, which controls the leading WKB
damping rate. Equivalently, since $H(r_c)=A(r_c)/r_c^2$,
\begin{equation}
 \lambda_c=\sqrt{-\frac{B(r_c)r_c^2H''(r_c)}{2}}.
 \label{eq:eikonal-lambda-simplified}
\end{equation}
The scalar mass $\mu$ contributes only to the subleading $O(L^{-1})$ correction
when $\mu$ is kept fixed as $L\to\infty$; the leading eikonal frequency is
controlled by the unstable null circular orbit of the corresponding black-hole
metric.

For the BK--1 black hole we keep the event-horizon radius
$r_h=2M$ explicit,

\begin{equation}
 A_1(r)=B_1(r)=\left(1-\frac{r_h}{r}\right)^2,
 \qquad
 H_1(r)=\frac{(r-r_h)^2}{r^4}.
\end{equation}

Here $A_1(r)$ and $B_1(r)$ are the BK--1 metric functions, and
$H_1(r)$ is the corresponding eikonal potential function $A_1(r)/r^2$. The
peak condition is

\begin{equation}
 H_1'(r)=\frac{2(r-r_h)(2r_h-r)}{r^5}=0,
\end{equation}

so the exterior maximum is at $r_c=2r_h$. At this point

\begin{align}
 H_1(2r_h)&=\frac1{16r_h^2},
 \quad
 H_1''(2r_h)=-\frac1{16r_h^4},\notag\\
 B_1(2r_h)&=\frac14.
\end{align}

Equations~\eqref{eq:eikonal-master-formula}--\eqref{eq:eikonal-lambda-simplified}
therefore give

\begin{equation}
 \omega_{\ell n}^{\rm BK1}=\frac{L}{4r_h}
 -i\left(n+\frac12\right)\frac{1}{4\sqrt2\,r_h}+\mathcal{O}(L^{-1})
 \label{eq:bk1-eikonal-qnm}
\end{equation}
for the extremal BK--1 black-hole metric. In this formula
$\omega_{\ell n}^{\rm BK1}$ denotes the BK--1 eikonal QNM frequency,
$L/(4r_h)$ is the leading real oscillation frequency, and
$1/(4\sqrt2\,r_h)$ is the BK--1 Lyapunov exponent. The normalized expressions
used previously are recovered by setting $r_h=1$.

For the BK--2 black-hole family, the horizon radius is the length
scale $a$ introduced in Eq.~\eqref{eq:bk2}; we denote it here by $r_h=a$ and keep
it explicit. The metric functions are

\begin{align}
 A_2(r)&=1-\frac{r_h^2}{r^2},\notag\\
 B_2(r)&=A_2(r)\left(1+\frac{C-1}{\sqrt{2r^2/r_h^2-1}}\right),
 \qquad C>0.
\end{align}

Here $A_2(r)$ and $B_2(r)$ are the BK--2 metric functions, and
$C$ is the dimensionless Bronnikov--Kim family parameter; the range $C>0$ is the
black-hole branch. The eikonal peak is independent of $C$, because

\begin{equation}
 H_2(r)=\frac{A_2(r)}{r^2}=\frac{r^2-r_h^2}{r^4},
 \quad
 H_2'(r)=\frac{2(2r_h^2-r^2)}{r^5}.
\end{equation}

Here $H_2(r)$ is the BK--2 eikonal potential function. Thus

\begin{equation}
 r_c=\sqrt2\,r_h,
 \quad
 H_2(r_c)=\frac{1}{4r_h^2},
 \quad
 H_2''(r_c)=-\frac{1}{r_h^4}.
\end{equation}

Only the Lyapunov exponent depends on $C$ through $B_2(r_c)$:

\begin{equation}
 B_2(r_c)=\frac12\left(1+\frac{C-1}{\sqrt3}\right)
 =\frac{C+\sqrt3-1}{2\sqrt3}.
\end{equation}

The BK--2 eikonal spectrum is therefore

\begin{equation}
 \omega_{\ell n}^{\rm BK2}=\frac{L}{2r_h}
 -i\left(n+\frac12\right)
 \frac{1}{r_h}\sqrt{\frac{C+\sqrt3-1}{2\sqrt3}}
 +\mathcal{O}(L^{-1})
 \label{eq:bk2-eikonal-qnm}
\end{equation}
for $C>0$. Here $\omega_{\ell n}^{\rm BK2}$ denotes the BK--2
eikonal QNM frequency, $L/(2r_h)$ is the leading real oscillation frequency, and
the square-root factor divided by $r_h$ is the BK--2 Lyapunov exponent. In
particular, the black hole
used in the continued-fraction scan, $C=1$, has

\begin{equation}
 \omega_{\ell n}^{\rm BK2,\,C=1}=\frac{L}{2r_h}
 -i\left(n+\frac12\right)\frac{1}{\sqrt2\,r_h}+\mathcal{O}(L^{-1}).
\end{equation}

In this last expression $\omega_{\ell n}^{\rm BK2,\,C=1}$ is the
eikonal QNM frequency of the BK--2 black hole
with $C=1$. Setting $r_h=1$ gives the dimensionless form used in
the numerical tables.

These eikonal expressions provide the large-$\ell$ black-hole check for the
single-peak barrier at fixed $\mu$. They complement the low-multipole massive
mode tracking performed above, where the quasi-resonant branches are followed
numerically. On the wormhole side, the corresponding high-frequency problem is a
two-ended scattering problem with
smooth passage through a regular throat and, in general, a two-barrier
resonant cavity; its eikonal treatment must therefore be formulated
as a separate two-ended resonance condition. Using the generic analytic
approach suggested in \cite{Konoplya:2023moy}, it is possible to extend the
above formulas beyond eikonal approximation.

Notice that the correspondence between the null geodesics and eikonal quasinormal modes are not guaranteed for all configurations. As was shown in \cite{Bolokhov:2023dxq,Konoplya:2022gjp,Konoplya:2017wot}, there are counterexamples. In particular, in some theories with higher curvature corrections, the eikonal instability may develop \cite{Konoplya:2017zwo,Dotti:2005sq,Gleiser:2005ra,Cuyubamba:2016cug}, break down the centrifugal term and consequently the correspondence. In our case the centrifugal term has the standard form $\sim A(r) \ell (\ell+1) r^{-2}$ and the correspondence is valid. For wormhole case, however, the correspondence will not work because of the double-well shape of the potential.


\section{Conclusion}

We have studied massive scalar quasinormal modes of Bronnikov--Kim brane-world
black holes and wormholes. The black hole
of the second Bronnikov--Kim black-hole family provides the main black-hole result: its fundamental branch,
several monopole overtones, and the dipole fundamental branch all approach
quasi-resonant endpoints. Their damping becomes very small, but the oscillation
frequency does not collapse to zero. This gives a clear counterexample to the
idea that the vanishing-real-frequency behaviour reported for the CFM
brane-world geometry \cite{Lutfuoglu:2026cfm} may be universal among brane-world black holes.

The wormhole calculations point in the same physical direction. Representative
Bronnikov--Kim throats also develop longer-lived massive scalar modes as the
field mass is increased. In the extended BK--2 wormhole scan, the damping of
the even-symmetry-sector fundamental mode
decreases by almost a factor of five compared with the massless case, while the
oscillation frequency remains finite.
Together with the BK--1 sample, this shows that long-lived massive-field ringing
is not tied to the presence of an event horizon, but can also occur on the
wormhole side of the Bronnikov--Kim parameter space.

The technical part of the analysis supports this conclusion in two independent
ways. For black holes, the Bronnikov--Kim--2 wave equation can be rewritten in
a rational form, so the Leaver continued-fraction method can be applied without
Taylor expanding the square root in the metric. For wormholes, a
one-exterior symmetry-reduced Chebyshev pseudospectral calculation gives a
complementary check of the same massive-field tendency. The positive-parameter pseudospectral tests further
confirm that the damping decreases with field mass while the real oscillation
frequency stays finite over the stable range of the finite-matrix calculation.
The eikonal WKB formulas add an independent high-multipole check: they
recover finite real frequencies governed by the null circular orbit and damping
rates governed by its Lyapunov exponent.

The broader lesson is that the massive scalar spectrum is sensitive to the
detailed brane-world geometry, not only to the fact that the spacetime is a
brane-world solution. The CFM zero-real-frequency phenomenon remains possible,
but it is not a generic feature of the Bronnikov--Kim families studied here.
What appears robust in these examples is the emergence of long-lived
massive-field ringing in both black holes and wormholes.

The precise
numerical QNMs obtained here can also be used to estimate grey-body
factors (GBFs). A recently proposed QNM--GBF correspondence relates the
transmission probability of a black-hole barrier to the fundamental QNM
frequency~\cite{Konoplya:2024gbf,Konoplya:2024vuj,Malik:2024cgb}. In the notation of Ref.~\cite{Konoplya:2024gbf},
the main relation, Eq.~(3.5) of that work, is
\begin{equation}
\begin{split}
 \Gamma_\ell(\Omega)\equiv |T|^2
 &=\left(1+\exp\left[2\pi
 \frac{\Omega^2-\operatorname{Re}(\omega_0)^2}
 {4\operatorname{Re}(\omega_0)\operatorname{Im}(\omega_0)}\right]\right)^{-1}\\
 &\quad +\mathcal{O}(\ell^{-1}).
\end{split}
 \label{eq:qnm-gbf-correspondence}
\end{equation}
Here $\Omega$ is the real scattering frequency, $\omega_0$ is the corresponding
fundamental QNM frequency, and $\Gamma_\ell$ is the partial grey-body factor.
Thus the high-accuracy Bronnikov--Kim QNM data provide a direct route to
estimating black-hole transmission probabilities and absorption observables. For
the wormhole side, an analogous use of the spectrum would require a two-ended
scattering formulation, but the correspondence takes place for wormholes \cite{Bolokhov:2024otn}.  However, the double-well structure makes the correspondence a poor tool for estimating grey-body factors. Future work could map the full dependence on the Bronnikov--Kim
parameter, extend the higher-multipole survey, and perform a dedicated two-ended
mode-tracking analysis of the wormhole resonance spectrum.

\begin{acknowledgments}
The author would like to thank Kirill Bornnikov and Sergei Bolokhov for useful discussions.
\end{acknowledgments}

\clearpage
\bibliographystyle{apsrev4-1}
\bibliography{references}

\end{document}